\documentclass[prb,preprint]{revtex4-1} 

\usepackage{array}
\usepackage{booktabs}
\usepackage{pifont}
\usepackage{mathtools}
\usepackage{amsfonts}
\usepackage{subcaption}
\usepackage{graphicx}
\usepackage{siunitx}
\usepackage[section]{placeins}
\usepackage[export]{adjustbox}

\newcommand*\rot{\rotatebox{90}}
\newcommand*\OK{\ding{51}}

\begin{document}

\title{Multiplicity counting using organic scintillators to distinguish neutron sources: An advanced teaching laboratory}

\author{Flynn B. Darby}
\email{fdarby@umich.edu} 
\author{Michael Y. Hua}
\author{Oskari V. Pakari}
\author{Shaun D. Clarke}
\author{Sara A. Pozzi}
\affiliation{Department of Nuclear Engineering and Radiological Sciences, University of Michigan, Ann Arbor, MI 48109}

\date{\today}

\newpage

\begin{abstract}

In this advanced instructional laboratory, students explore complex detection systems and nondestructive assay techniques used in the field of nuclear physics. 
After setting up and calibrating a neutron detection system, students carry out timing and energy deposition analyses of radiation signals. 
Through the timing of prompt fission neutron signals, multiplicity counting is used to carry out a special nuclear material (SNM) nondestructive assay. 
Our experimental setup is comprised of eight trans-stilbene organic scintillation detectors in a well-counter configuration, and measurements are taken on a spontaneous fission source as well as two ($\alpha$,n) sources. 
By comparing each source's measured multiplicity distribution, the resulting measurements of the ($\alpha$,n) sources can be distinguished from that of the spontaneous fission source. 
Such comparisons prevent the spoofing, i.e., intentional imitation, of a fission source by an ($\alpha$,n) neutron source. 
This instructional laboratory is designed for nuclear engineering and physics students interested in organic scintillators, neutron sources, and nonproliferation radiation measurement techniques.

\end{abstract}

\maketitle


\section{Introduction} 


In this advanced instructional laboratory, students learn the fundamentals of a current International Atomic Energy Agency (IAEA) nondestructive assay technique. 
To carry out this study, students measure neutron sources with multiple detectors and then learn how to analyze the resulting timing and energy data in order to identify the characteristic signatures of varying neutron sources. 
Such detection and characterization of neutron sources is a crucial goal in monitoring nuclear nonproliferation.\cite{runkle_securing_2010,doyle_nuclear_2019}
Special nuclear material (SNM) such as plutonium and enriched uranium is identifiable through the detection of neutrons and photons emitted from spontaneous and induced fission events. 
However, when only the single-neutron count rates from fission are considered, the detection signals of SNM can be spoofed, i.e., intentionally imitated, by substitution of non-fissioning neutron sources.
An undetected spoof can lead to an unknown SNM diversion, i.e., concealing SNM from its declared location, which can in turn lead to global security risks.

Nuclear fusion and ($\alpha$,n) reactions mainly emit single neutrons,\cite{petrescu_basic_2017,lorch_neutron_1973} while nuclear fission reactions emit multiplets of neutrons and gamma rays for each reaction and the shape of the multiplicity distribution is isotope-dependent.\cite{fraser_nuclear_1966} 
Therefore, fission sources may be distinguished from single-neutron-emitting radiation sources. 
In addition, the fissioning isotope may be inferred from the detected neutron multiplicity distributions using multiplicity counting.

Multiplicity counting is a method in which detected neutron arrival times are analyzed with sequential timing gates, in which counts are recorded in each timing gate. 
The resulting count distribution is converted to effective multiplet count rates to determine the mass of SNM.
The SNM sample's total mass and isotopic masses are calculated from the measured multiplets if isotopic ratios are known from complementary gamma-ray spectroscopy.
The IAEA uses neutron multiplicity counting measurements as a nondestructive assay method to verify plutonium and uranium amounts in samples. 
Conventionally, polyethylene-moderated $^3$He detectors are used to acquire the timing signals for these measurements.\cite{macklin_neutron_1972,langner_neutron_1991,krick_thermal-neutron_1992} 
The polyethylene moderation scatters and slows down incident neutrons before capture in the $^3$He gas, hence increasing the likelihood of a $^3$He neutron-capture reaction. 
These detectors are arranged in a so-called ``well-counter'' configuration, where a cylindrical array of detectors is oriented about a central fission source.

Organic scintillators detect neutrons by elastic scattering without the need for moderation and, with their nanosecond timing precision, improve the ratio of correlated-to-uncorrelated signals. 
When compared to moderated thermal neutron detection methods, these attributes can significantly decrease the measurement time required for a low-uncertainty assay. \cite{di_fulvio_passive_2017,hua_measured_2020} 
The main drawback of an organic scintillator well-counter is particle cross-talk, i.e., the scattering of one particle into multiple detectors, which causes a spurious correlated signal. 
Recently, theoretical adjustments for neutron cross-talk\cite{shin_neutron_2015} enabled the use of organic scintillators for multiplicity counting applications.\cite{shin_neutron_2017} 
Additionally, organic scintillators are sensitive to both neutron and photon radiation and can discriminate between the two types.\cite{brooks_pulse_1960,winyard_pulse_1972} 
This dual-particle sensitivity has the potential to further improve the precision of correlated fission radiation measurements through increased statistics as well as add additional signatures for fission source analysis.\cite{pazsit_combined_2009,hua_gamma-ray_2019,pakari_high_2022}

In this work, we take measurements on one $^{252}$Cf fission source and two ($\alpha$,n) neutron sources, calculate discriminated neutron and photon energy deposition spectra, and apply mixed multiplicity counting using a well counter \cite{di_fulvio_fast_2015,shin_fast_2016,shin_fast-neutron_2017,di_fulvio_passive_2017,shin_neutron-neutron_2019} of eight trans-stilbene \cite{zaitseva_scintillation_2015} organic scintillators.
We show that this approach can be used to distinguish between the two types of neutron sources.
We use commercially available equipment for detector operation and readout and discuss how to conduct the experiment and analyze the data.


\section{Background}

The three neutron sources utilized in this laboratory are a $^{252}$Cf spontaneous fission source, a $^{239}$Pu-Be (PuBe) ($\alpha$,n) source, and a $^{241}$Am-Li ($\alpha$,n) source. 
The $^{252}$Cf spontaneous fission source is used as a stand-in for fissioning special nuclear material (SNM), i.e., plutonium and uranium. 
The $^{252}$Cf has a much higher multiplicity than plutonium and uranium and $^{252}$Cf only emits neutrons by spontaneous fission.
For this reason, $^{252}$Cf is easily distinguishable from ($\alpha$,n) sources when the multiplicity of each source is considered.
PuBe is predominantly ($\alpha$,n), but does include trace amounts of plutonium (SNM) with an associated multiplicity.
AmLi is an ($\alpha$,n) neutron source without any SNM.

The ($\alpha$,n) neutron sources used in this laboratory are $^{239}$Pu-Be (PuBe) and $^{241}$Am-Li (AmLi). 
The ($\alpha$,n) reactions for PuBe and AmLi have average neutron energies of approximately 4.2 and 0.5 MeV and endpoint energies of approximately 10.8 and 1.2 MeV, respectively. \cite{geiger_radioactive_1975,weaver_neutron_1982} 
The gamma-ray spectra of both neutron sources are complex, but both include photopeaks due to the de-excitation of the alpha-collided nucleus --  $^9$Be($\alpha$,n)$^{12}$C$^*$ emits a 4.4 MeV gamma ray and $^7$Li($\alpha$,$\alpha^\prime$)$^7$Li$^*$ emits a 0.5 MeV gamma ray.\cite{vega-carrillo_neutron_2002,moore_configuration_2021}

\begin{figure}[h]
    \centering
    \includegraphics[trim= 30 180 60 200,clip,width=0.6\linewidth]{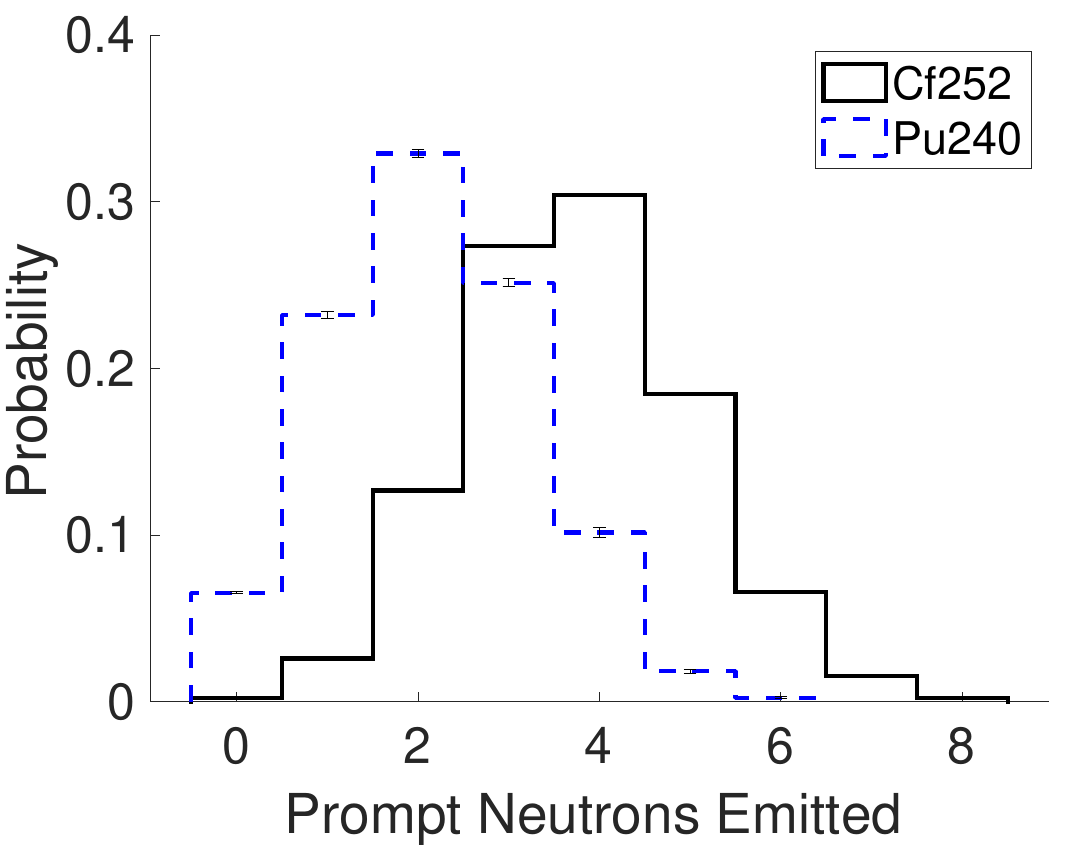}
    \caption{(color online) The prompt neutron multiplicity distribution of $^{252}$Cf and $^{240}$Pu spontaneous fission.\cite{boldeman_prompt_1985} The associated uncertainties are within the lines for both distributions.}
    \label{fig:boldhinesmults}
\end{figure}

The $^{252}$Cf source emits neutrons through spontaneous fission. 
Although the majority of decays are alpha emissions (97\%), the $^{252}$Cf is not surrounded by a low-Z matrix that would lead to ($\alpha$,n) reactions. 
The prompt neutron emission energy follows a Watt spectrum with an average energy of approximately 2.1 MeV. \cite{smith_spontaneous_1957} 
For each spontaneous fission event, $^{252}$Cf may emit 0 to 8 (or more) neutrons and 0 to 20 (or more) gamma rays.  
On average, $^{252}$Cf emits 3.76 neutrons and about 8 gamma rays per spontaneous fission.\cite{boldeman_prompt_1985,valentine_evaluation_2001} 
Trace amounts of $^{240}$Pu in the PuBe source also emit neutrons through spontaneous fission with an average energy of about 2 MeV.\cite{vladuca_improved_2001}
Each $^{240}$Pu spontaneous fission event emits 0 to 6 (or more) neutrons and 2.16 neutrons on average. \cite{boldeman_prompt_1985} 
According to Boldeman and Hines, the full neutron multiplicity distributions for both $^{252}$Cf and $^{240}$Pu  are plotted in Fig.~\ref{fig:boldhinesmults}. 
$^{239}$Pu may also fission from neutron-induced fission events and the multiplicity is dependent on the incident neutron energy.\cite{marini_energy_2022} 
Additionally, $^9$Be has an (n,2n) reaction cross-section similar to its ($\alpha$,n) cross-section. 
Both cross sections are between 0.1-1 b at incident particle energies of 4-10 MeV.\cite{soppera_JANIS_2014} 
The (n,2n) reaction has the potential to cause non-fission coincident neutron detection events, but this effect will not contribute to true detected fission multiplets greater than two.

The PuBe ($\alpha$,n) source is treated as predominantly ($\alpha$,n) in this laboratory. 
We do not have access to the exact elemental ratios and isotopic ratios from the manufacturer, but previous work has shown the atomic ratio of beryllium-to-plutonium atoms in the matrix is typically 13:1.\cite{bagi_neutron_2016} 
Additionally, the likelihood of PuBe ($\alpha$,n) neutron production is estimated to significantly outweigh correlated neutron emissions.

To measure the radiation from these neutron sources, organic scintillators coupled with photomultipler tubes (PMTs) will be utilized. 
The organic scintillators detect neutrons mainly through elastic scattering on protons and photons through Compton scattering on electrons, both in the hydrocarbon active volume. 
The two types of radiation are distinguishable due to the ratio of prompt and delayed scintillation light. 
Most of the scintillation light comes from prompt fluorescence (few nanoseconds), but the higher stopping power ($dE/dx$) of a proton in the active volume leads to an increased density of triplet states that de-excite through delayed fluorescence (hundreds of nanoseconds).\cite{birks_theory_1964}

\begin{figure}[h]
    \centering
    \begin{subfigure}{0.49\textwidth}
        \includegraphics[trim= 30 180 60 200,clip,width=\textwidth]{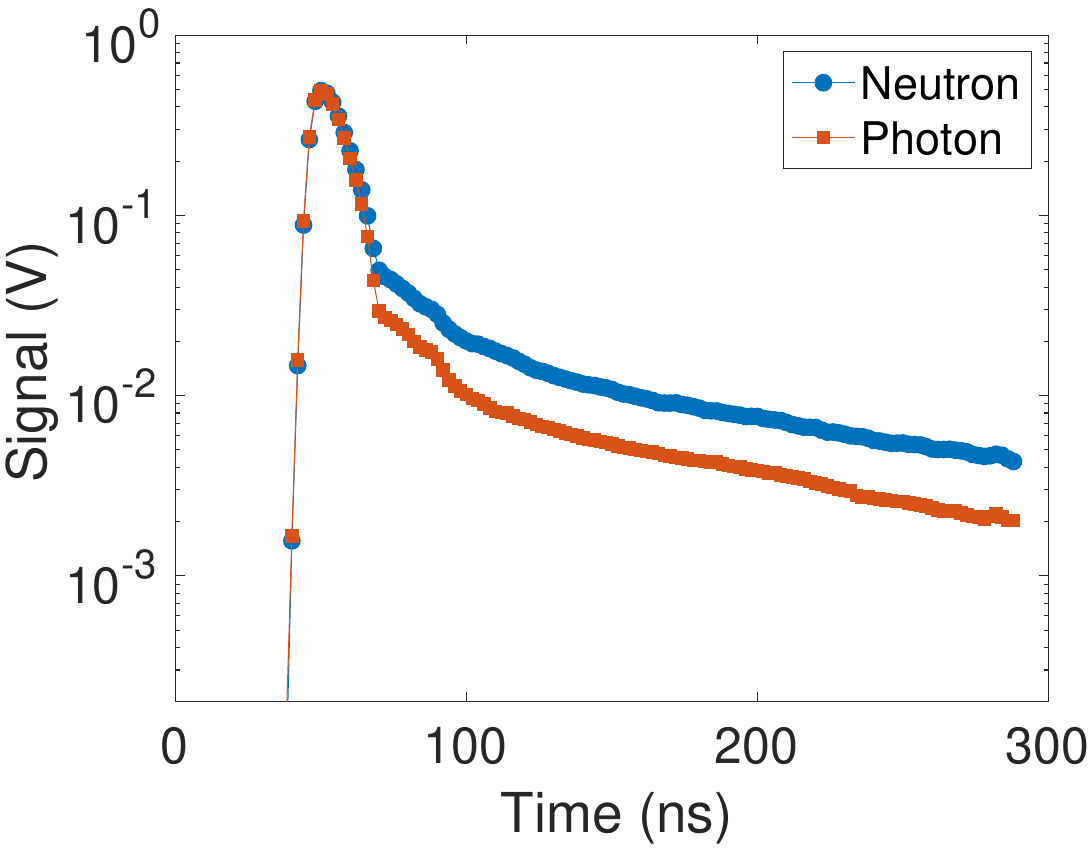}
        \caption{}
        \label{fig:averagePH}
    \end{subfigure}
    \begin{subfigure}{0.49\textwidth}
        \includegraphics[trim= 30 180 60 200,clip,width=\textwidth]{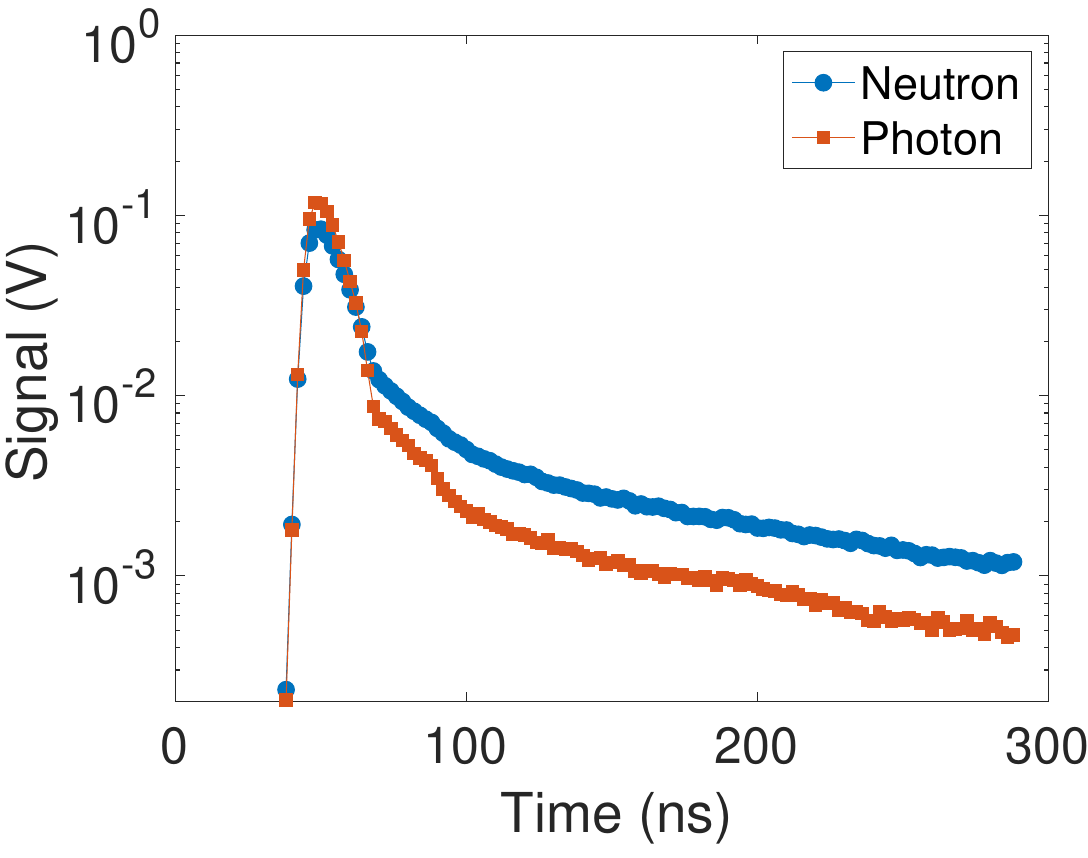}
        \caption{}
        \label{fig:averagePI}
    \end{subfigure}
    \caption{(color online) The mean of 100 averaged neutron or 100 averaged photon pulses of (a) the same $0.50 \pm 0.01$ V pulse amplitude and (b) the same $2.00 \pm 0.01$ V-ns pulse integral. Negative data and near-zero data (due to baseline subtraction) are not shown on a logarithmic scale.}
    \label{fig:average_pulses}
\end{figure}

Because of the delayed fluorescence, the pulse die-away time for neutron scatters will be longer than that of photon scatters of equivalent light output, which may be quantified by either pulse height or pulse integral.
Thus, we can discriminate between the two types of interactions with charge-integration pulse-shape discrimination (PSD).\cite{polack_algorithm_2015,brooks_pulse_1960} 
Figure~\ref{fig:average_pulses} displays the clear difference in the tail region of neutron and photon pulses.
For a neutron and photon signal of equivalent pulse height, the tail region of the neutron signal will have a higher integral value. 
The same remains true for a neutron and photon signal of equivalent pulse integral.
This characteristic highlights the significant amount of delayed fluorescence associated with neutron signals relative to photon signals.

To interpret the energy deposition of particle scatters in organic scintillators, a gamma-ray source of known mono-energetic energy is required to calibrate each detector. 
Hence, a $^{137}$Cs source is used to calibrate each detector using its mono-energetic 662 keV gamma ray. 
The Compton edge energy (CE) is calculated as 

\begin{equation}
    \text{CE} = E_\gamma \left(1-\frac{1}{1+\frac{2E_\gamma}{m_ec^2}}\right),
    \label{eq:CE}
\end{equation}

\noindent where $E_\gamma$ is the initial energy (662 keV) of the emitted gamma ray and $m_ec^2$ is the rest mass of an electron (511 keV). 
Thus, a Compton edge of 478 keV is used to calibrate the light output of each detector assuming zero energy deposition does not result in a pulse.

\begin{figure}[h]
    \centering
    \includegraphics[trim= 30 180 60 200,clip,width=0.6\linewidth]{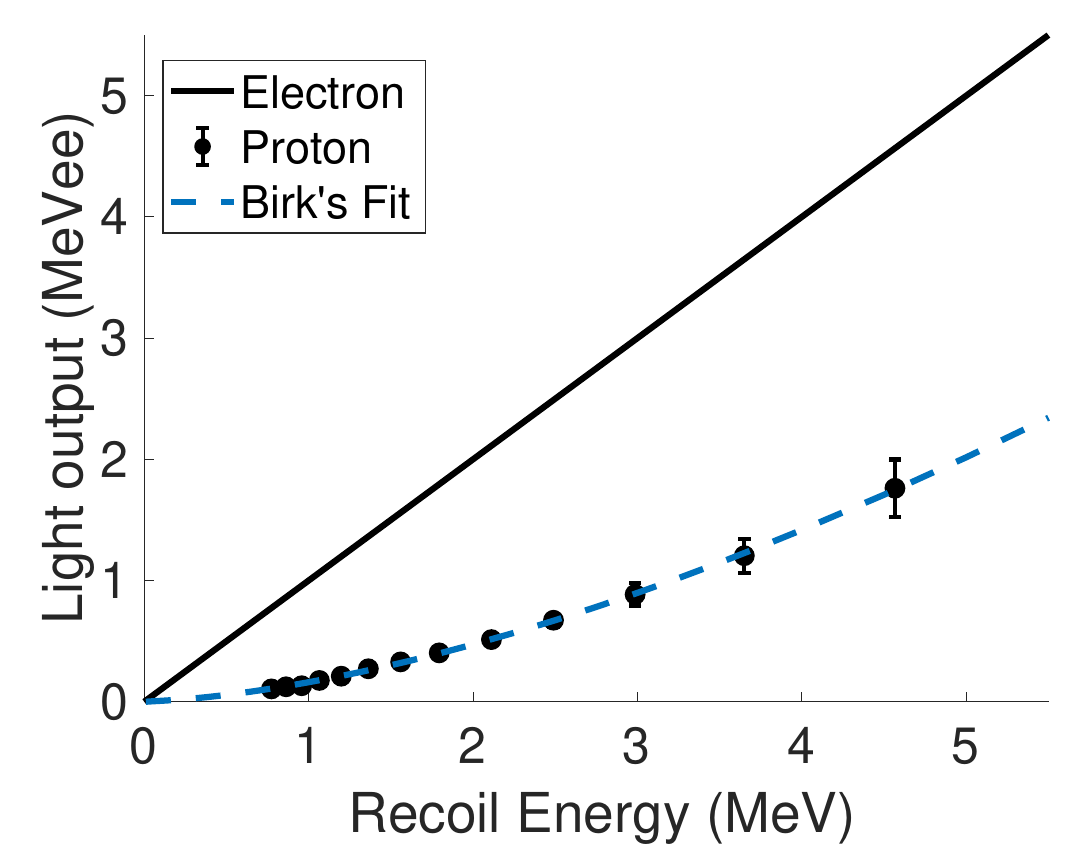}
    \caption{(color online) Proton light output relative to electron light output calculated and fit for our trans-stilbene detector by Shin.\cite{shin_measured_2019}}
    \label{fig:neutronLO}
\end{figure}

The different interaction types and fluorescence pathways also lead to quenched light output for neutrons.\cite{craun_analysis_1970} 
Figure~\ref{fig:neutronLO} shows the relative light output due to protons (scattered by neutrons) and electrons (scattered by photons) in our trans-stilbene detector, as measured by Shin.\cite{shin_measured_2019} 
A neutron scatter of 2.00 MeV energy deposited on a proton results in a light output of 0.47 MeV-electron-equivalent (MeVee) according to the Birk's fit: 

\begin{equation}
    \text{L}(E) = \int \frac{a}{1+b\frac{dE}{dx}} dE,
    \label{eq:Birks}
\end{equation}

\noindent where $\text{L}(E)$ is the light output due to a proton recoil, $\frac{dE}{dx}$ is the energy-dependent stopping power of the proton, and $a$ and $b$ are constants calculated by fitting data from a light output experiment (see Fig.~\ref{fig:neutronLO}). 
Shin calculated $a=1.694$ and $b=29.05$ for the stilbene detectors used in this work. 
The Birk's fit is commonly accepted for neutron light output as a function of deposited energy.\cite{norsworthy_evaluation_2017}

Lastly, the neutron scatter nature of organic scintillators leaves an array susceptible to cross talk events. 
Cross talk occurs when one particle scatters in one detector followed by the scattering of that same particle in another detector with both scatter events exceeding the detection energy deposition threshold.
This sequence would lead to spurious coincident events in an organic scintillator array. 
Cross talk will be acknowledged in this work, but additional simulations outside the scope of this paper are required to fully account for cross talk in organic scintillator systems.


\section{Experimental Setup}

Each neutron source was shielded with 2.16 cm of annular lead shielding during measurement.
When neutron sources were not in use, each source was stored in a large metal containment filled with low-Z, paraffin wax in a source room attached to the laboratory with a unique key entry.
The $^{137}$Cs button source is stored in a small lead pig, a cylindrical lead containment with a lead cap that forms a full shell around the source, when not in use in the same source room.

\begin{figure}[h!]
    \centering
     \includegraphics[width=0.8\linewidth]{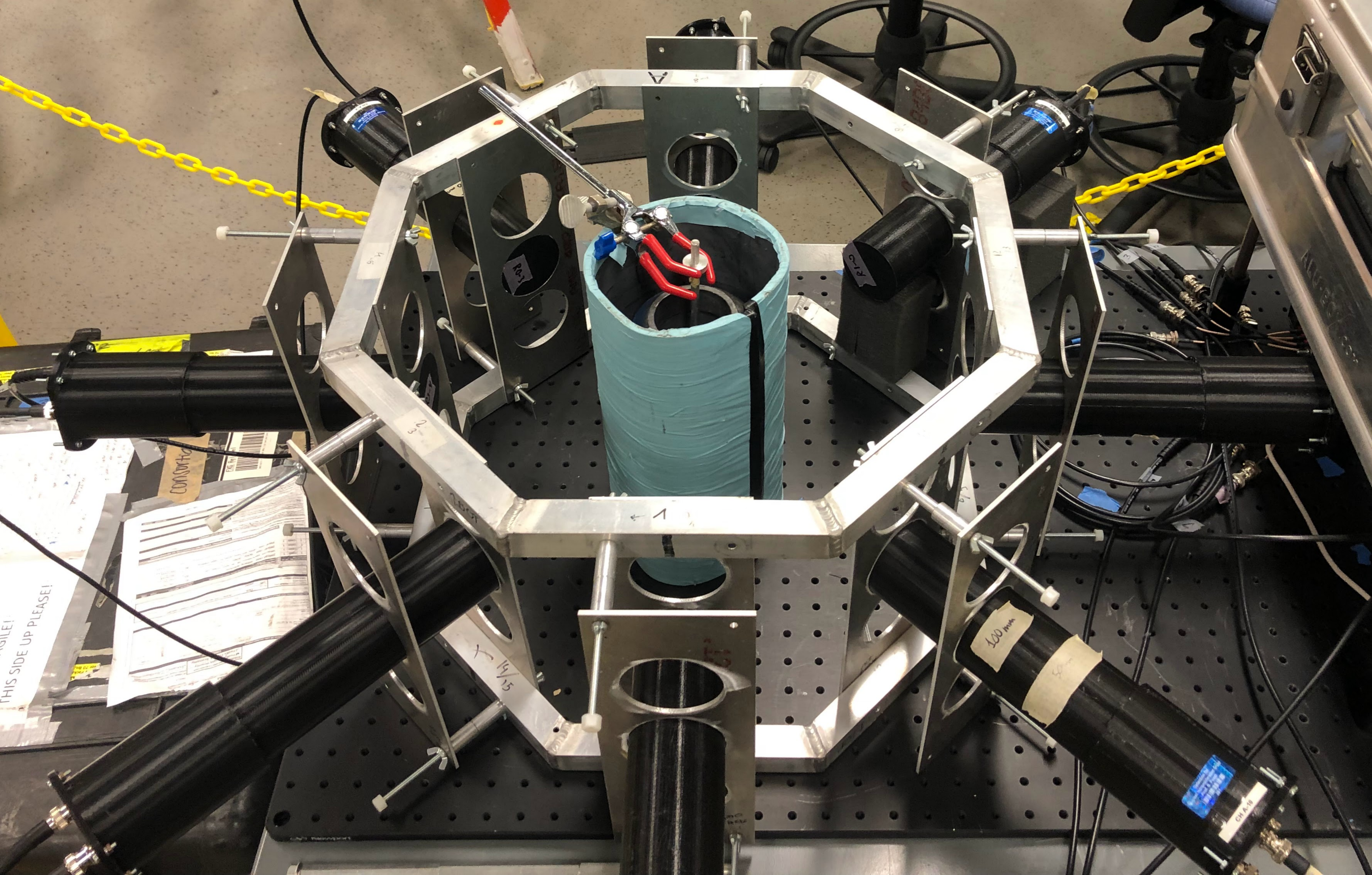}
    \caption{Eight trans-stilbene detectors in the well-counter frame. The detectors' vertical positions and faces are centered about each source and the face of each detector is 22 cm from the source center. In this setup, our AmLi source is shielded by 2.16 cm of annular lead shielding.}
    \label{fig:wellcounterAmLi}
\end{figure}

\begin{figure}[h!]
    \centering
    \includegraphics[width=0.8\linewidth]{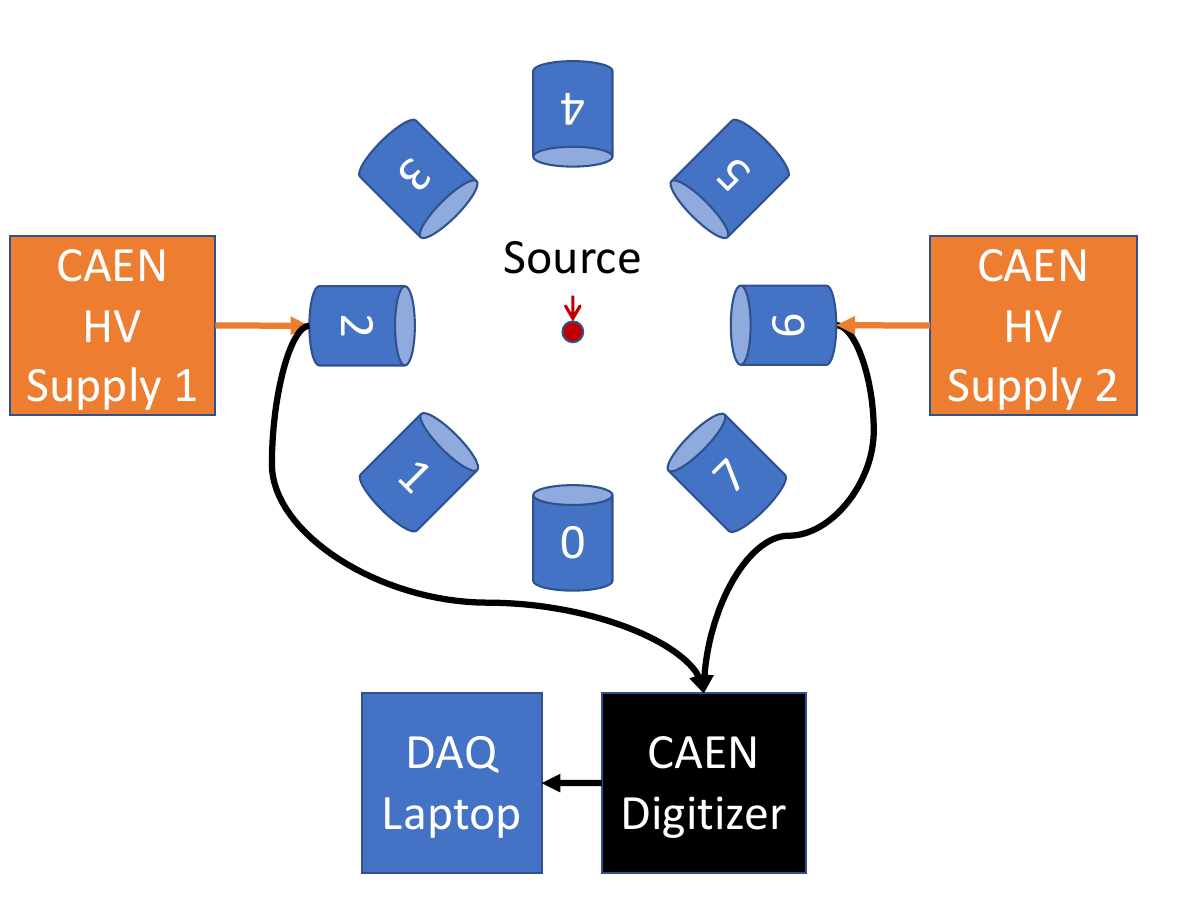}
    \caption{(color online) An axial diagram of the well counter connected to operating and readout electronics. The radiation source in the diagram is positioned at the center of the detector array. The SHV cables are orange and the BNC cables are black. There is a pair of SHV and BNC cables for each detector, but only two pairs are shown here for simplicity.}
    \label{fig:bd}
\end{figure}

We operated our detection system with a laptop using CoMPASS data acquisition software, a CAEN DT5730S 500-MHz, 14-bit digitizer, and two CAEN DT1470ET high voltage units to measure the sources in Table~\ref{tab:sources}. 
The eight trans-stilbene organic scintillators were arranged in a circular, well-counter frame, evenly spaced about the circumference as shown in Fig.~\ref{fig:wellcounterAmLi}. 
Each detector is separated by 45 degrees and the detector face and vertical positions are centered about the source. 
The center of each source is 22 cm from the face of each detector. 
Each detector is a housed assembly of a 5.08 cm diameter by 5.08 cm length trans-stilbene crystal coupled to a 5.08 cm face photomultiplier tube with BNC readout and SHV input inside a plastic, 3D-printed casing. 
Figure~\ref{fig:bd} provides a diagrammatic representation of the connections as well as the central source position. 
The digitizer connections required BNC-MCX adapters. 
The analog signals from the detectors are read by the digitizer on a 2 V dynamic range and output to the laptop if the signal meets acquisition requirements. 
Throughout the measurements, a Ludlum Model 30 neutron dose meter and a Fluke 451P-RYR Ion Chamber were used to monitor neutron and gamma dose rates, respectively.

\begin{table}[h!]
\caption{The sealed sources used in this experiment.}
\label{tab:sources}
\begin{tabular}{l|l|l|l|l|l}
 & &  & \multicolumn{1}{c|}{Dated} & \multicolumn{1}{c|}{Activity} & \multicolumn{1}{c}{Neutron}                    \\
\multicolumn{1}{c|}{Source} & \multicolumn{1}{c|}{Maker} & \multicolumn{1}{c|}{Date} & \multicolumn{1}{c|}{Activity (Ci)} & \multicolumn{1}{c|}{when measured (Ci)} & \multicolumn{1}{c}{emission rate (n/s)}                    \\ \hline
$^{137}$Cs & EZ\footnote[1]{Eckert and Ziegler} & 10/15/16 &  $9.76\times10^{-5}$ & $8.53\times10^{-5}$ & N/A \\
$^{252}$Cf & FTC\footnote[2]{Frontier Technology Company} & 05/08/15 & $5.63\times10^{-3}$ &  $8.93\times10^{-4}$  & $3.83\times10^6$    \\
PuBe       & MRC\footnote[3]{Monsanto Research Corporation} & 09/30/63 &  $1.00$  & $9.98\times10^{-1}$ & $1.70\times10^6$              \\
AmLi       & Gt\footnote[4]{Gammatron} & 12/10/19 & 9.60 & 9.56  &  $5.74\times10^5$                         \\
\end{tabular}
\end{table}


\section{Procedure}


During all measurements, ALARA (as low as reasonably achievable) dose practices were applied. 
We used the minimum time and reasonable distance and shielding to minimize the radiation dose to participants. 
We shielded the AmLi source with lead to decrease the gamma dose and improve the neutron count ratio in our detectors. 
No more than three meters of separation of participants from the sources was required to reduce neutron dose below 0.5 mrem/hr and gamma dose below 0.1 mrem/hr measured by calibrated dose meters.

\begin{figure}[h!]
    \centering
    \includegraphics[trim= 30 180 60 200,clip,width=0.6\linewidth]{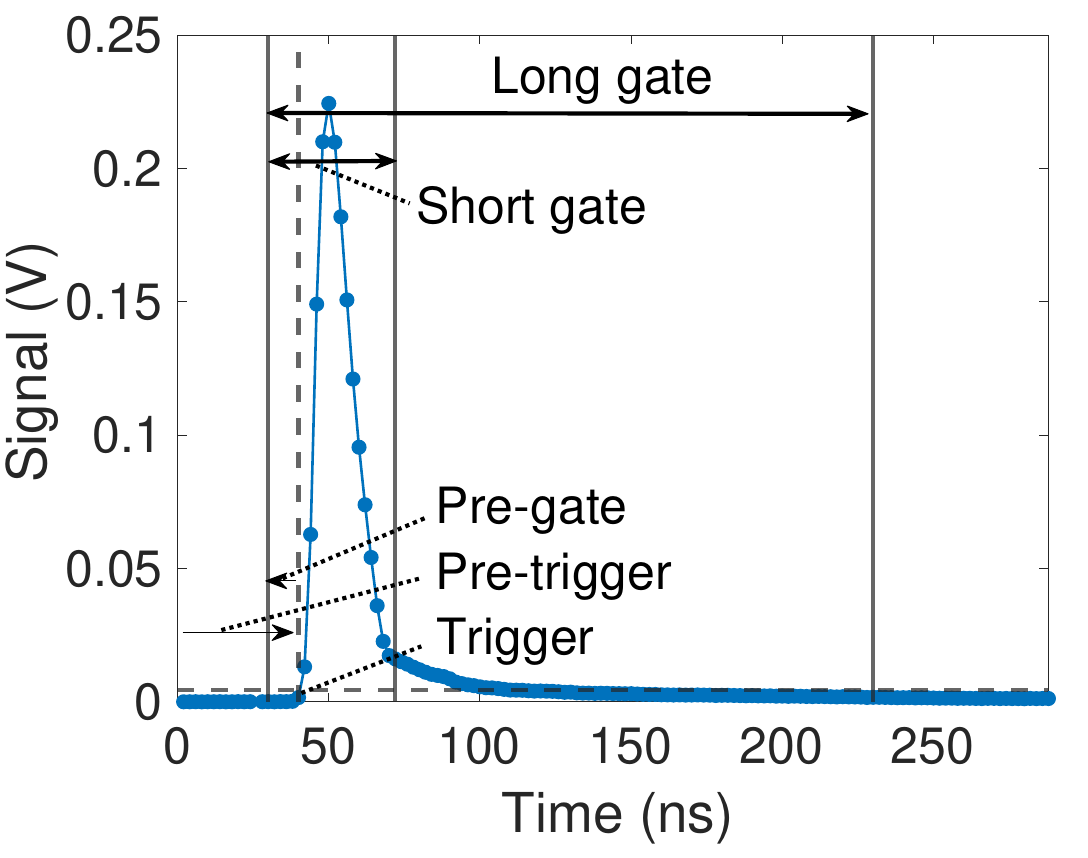}
    \caption{(color online) One hundred averaged pulses of $4.00\pm0.01$ V-ns integral with annotated pulse integral gate settings.}
    \label{fig:custom_PSD_diagram}
\end{figure}

We acquired pulse data with our digitizer controlled by a Dell laptop using the Windows 10 CAEN CoMPASS data acquisition software. 
The PSD and pulse window settings, based on previous work,\cite{shin_measured_2019} are as follows. 
The pulse acquisition window was set to 288 ns with a 200 ns pulse integral gate, 42 ns short gate, 10 ns pre-gate (or ``gate offset''), and 40 ns pre-trigger.
We triggered on the leading edge at a 35 LSB (least-significant bits) threshold above the baseline leading to a pulse height threshold of $(\si{35 }{\, \text{LSB}})\times(\frac{\si{2 }{\, \text{Vpp}}}{\si{2^{14}-1 }{\,  \text{LSB}}}) = 4.3 \, \text{mV}$. 
The baseline was dynamically calculated for each waveform using the first 16 samples (32 ns).
The digital scaling was set to 2.5 fC/(LSB $\times$ Vpp). 
Figure~\ref{fig:custom_PSD_diagram} provides a diagrammatic explanation of the gate settings. When a pulse exceeds the 4.3 mV threshold, the digitizer triggers and writes out a full waveform, as seen in Fig.~\ref{fig:custom_PSD_diagram}. 
The time stamp associated with the trigger is calculated by the digitizer as the linearly interpolated time between the first sample below the trigger threshold and the first sample above the trigger threshold.

\subsection{Light-output calibration}

To gain-match each of our detectors, we use adjustments in the high voltage applied to each detector in order to align the spectral channels to the Compton edge from the $^{137}$Cs 662 keV gamma ray. 
We measured the $^{137}$Cs source without lead shielding and adjusted the applied voltage on each detector to align the Compton edge of the pulse integral light-output spectrum for each channel within 1.95 $\pm$ 0.04 V-ns. 
An ideal pulse integral distribution exhibits a sharp drop-off after a peak at the Compton edge. 
Realistically, this sharp edge is distorted by detector energy resolution (in this case, 12.90 +/- 0.24\%),\cite{shin_fast-neutron_2019} leading the true location of the Compton edge to occur at a light output greater than the measured peak. 
The true location can be determined by performing an energy-time gated measurement.
Such a measurement has previously been completed and described by Shin,\cite{shin_fast-neutron_2019} and it was found that the true Compton edge appears at 75.4\% of the observed Compton maximum frequency corresponding to a higher pulse integral value than the observed Compton maximum. 
Once all channels were calibrated, we collected a final calibration measurement of five minutes.


\subsection{Neutron source measurements}

With our detectors calibrated, we measured the $^{252}$Cf source and the PuBe and AmLi ($\alpha$,n) sources at the center of our well counter shielded by 2.16 cm of annular lead shielding similar to Fig.~\ref{fig:wellcounterAmLi}. 
We measured the two strongest neutron sources, $^{252}$Cf and PuBe, for five minutes and the weakest neutron source, AmLi, for 30 minutes. 
Measurement times are dependent on source strength.
In this work, we have detected over one million neutrons for each source.


\section{Data processing}
\label{sec:datproc}

We acquired list format data of the channel, time of detection (trigger), long gate, and short gate for each detection on-the-fly with our digitizer. 
CoMPASS PSD settings allow on-the-fly acquisition of the long gate and short gate pulse integrals in analog-to-digital (ADC) digitizer units. 
To check the analog-to-digital scaling of our raw waveforms, we briefly collected full waveforms during calibration. 
For convenience as well as bandwidth restrictions of our data acquisition setup, only list format data was processed for the remaining measurements.
During data acquisition with the $^{252}$Cf source, we received a memory error when attempting to collect full waveforms at a data throughput rate of approximately 13.15 MB/s. 
This was likely due to the hard drive write speed of our laptop. 
By collecting in list format, our throughput was reduced to about 1.53 MB/s and the memory error did not occur.

\subsection{Light-output conversion factor}

\begin{figure}[h!]
    \centering
    \begin{subfigure}{0.49\textwidth}
        \includegraphics[trim= 30 180 60 200,clip,width=\textwidth]{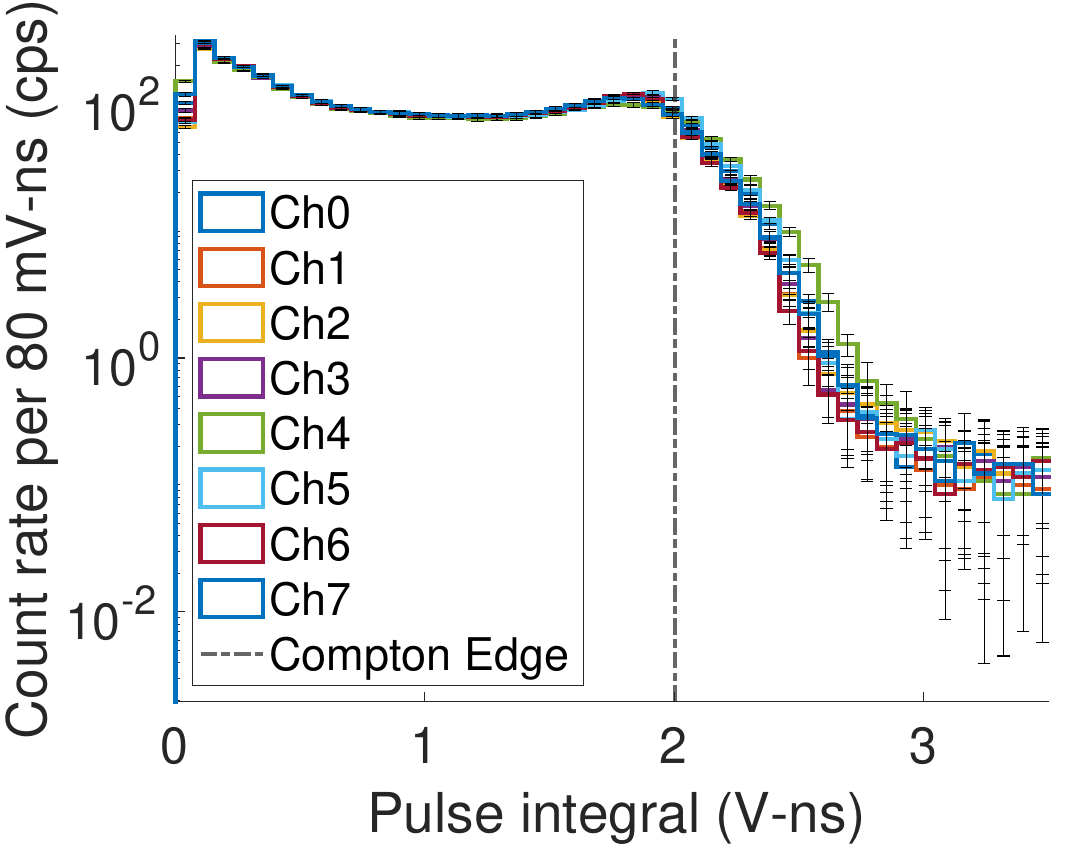}
        \caption{}
        \label{fig:calib_adc}
    \end{subfigure}
    \begin{subfigure}{0.49\textwidth}
        \includegraphics[trim= 30 180 60 200,clip,width=\textwidth]{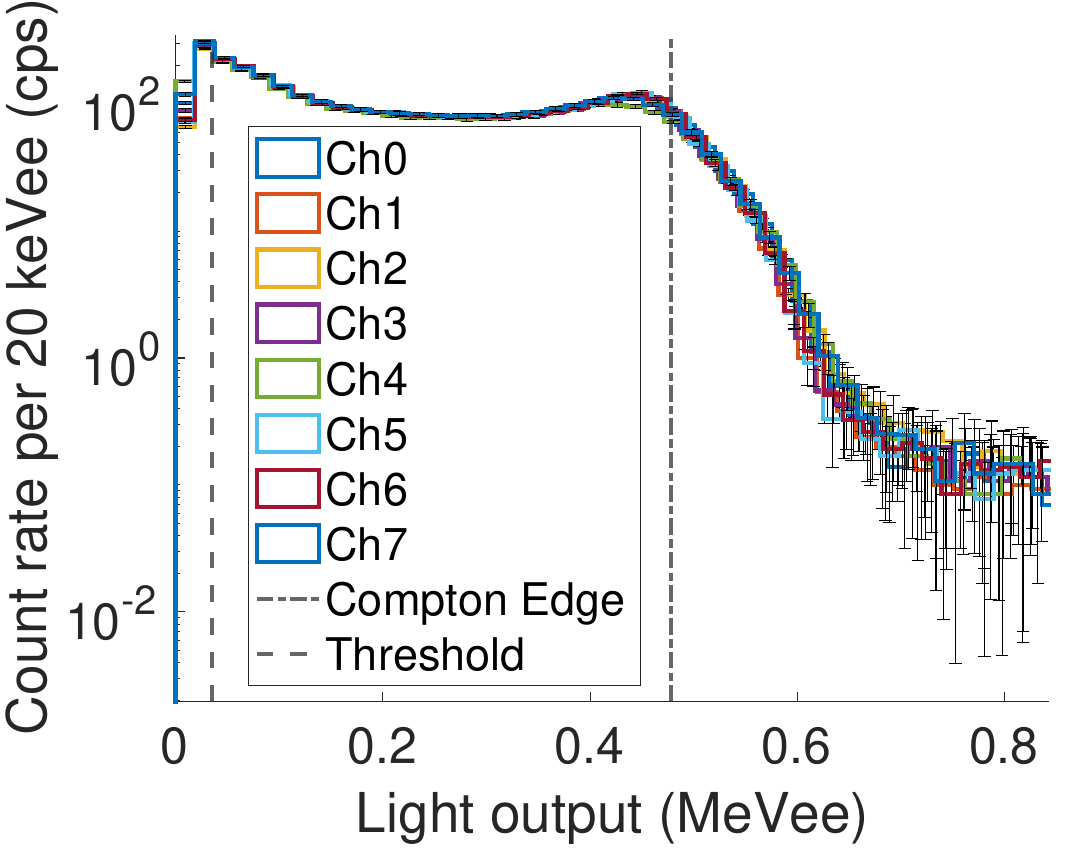}
        \caption{}
        \label{fig:calib_MeVee}
    \end{subfigure}
    \caption{(color online) Calibration of light-output for each channel from, (a) V-ns units to (b) electron-equivalent for each channel with a light-output conversion of 0.245 $\pm$ 0.004 MeVee/V-ns. We represent a post-processing 0.035 MeVee threshold that is applied with the vertical black line on the electron-equivalent spectrum.}
    \label{fig:calib}
\end{figure}

Using the long gate integral values from our list format data, we converted our calibrated $^{137}$Cs pulse-integral spectra to light-output spectra in Fig.~\ref{fig:calib}, taking our true Compton edge to be at 75.4\% of the peak amplitude due to the inherent detector light-output resolution.\cite{shin_fast-neutron_2019} We calculated a light-output conversion factor of $\frac{0.478 \text{ MeVee}}{1.95 \pm 0.03 \text{ V-ns}} = 0.245 \pm 0.004$ MeVee/V-ns, accounting for one sample standard deviation from averaging the eight channels.

\subsection{Pulse-shape discrimination (PSD)}

\begin{figure}[h!]
    \centering
    \begin{subfigure}{0.49\textwidth}
        \includegraphics[trim= 30 180 60 200,clip,width=\textwidth]{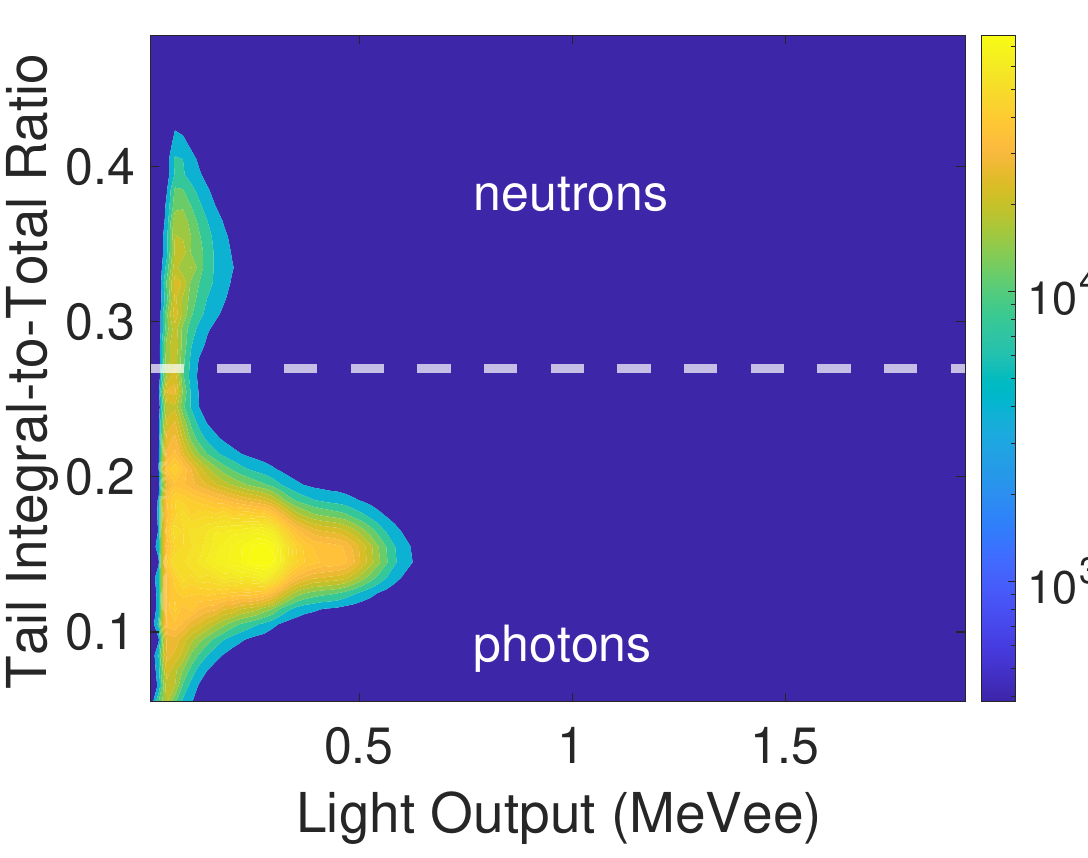}
        \caption{}
        \label{fig:PSD_AmLi}
    \end{subfigure}
    \begin{subfigure}{0.49\textwidth}
        \includegraphics[trim= 30 180 60 200,clip,width=\textwidth]{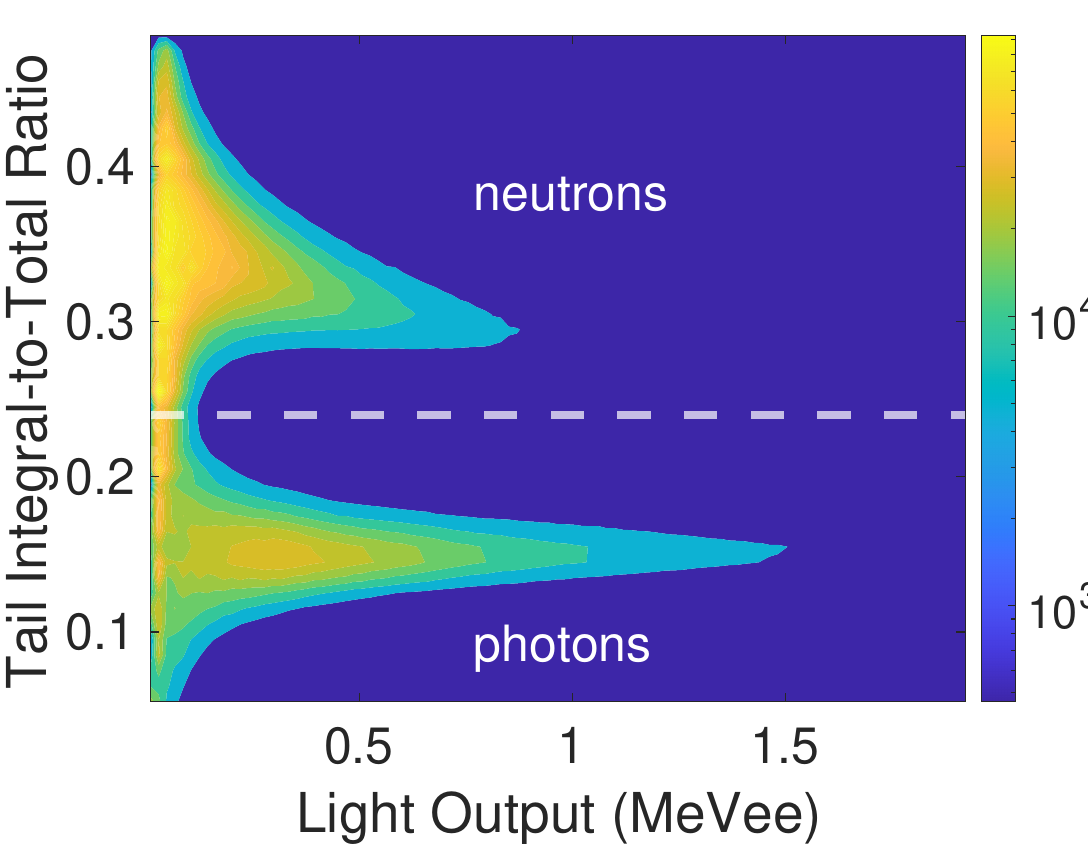}
        \caption{}
        \label{fig:PSD_Cf252}
    \end{subfigure}
    \begin{subfigure}{0.49\textwidth}
        \includegraphics[trim= 30 180 60 200,clip,width=\textwidth]{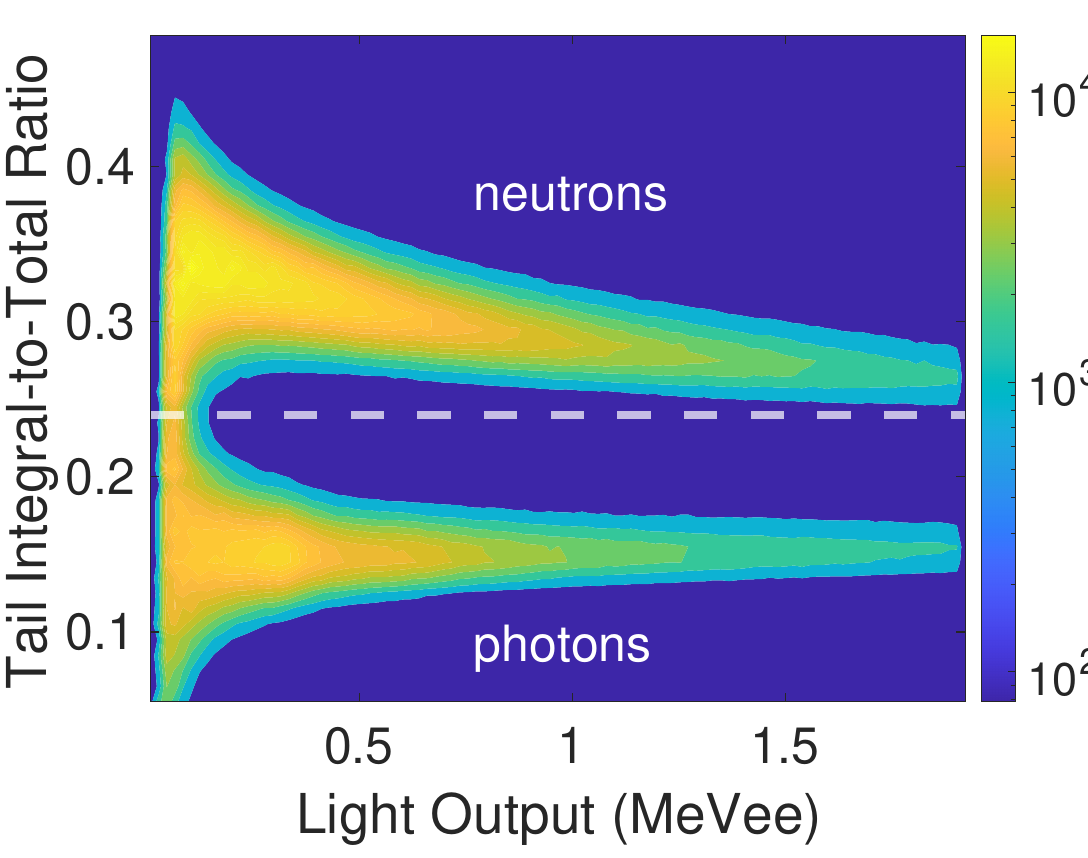}
        \caption{}
        \label{fig:PSD_PuBe}
    \end{subfigure}
    \caption{(color online) Pulse-shape discrimination plots combining the data from all detectors for the (a) AmLi, (b) $^{252}$Cf, and (c) PuBe measurements. The color bar units are in frequency (counts per second) per heatmap cell.}
    \label{fig:PSD}
\end{figure}

After calibrating each channel with the $^{137}$Cs source, we used a pulse-shape discrimination (PSD) parameter known as the tail-to-total integral ratio to discriminate the neutrons from photons for each neutron source.
The tail-to-total integral ratio (R) is:

\begin{equation}
    \label{eq:PSD}
    \text{R} = \frac{\text{LG}-\text{SG}}{\text{LG}} = \frac{\text{tail integral}}{\text{total integral}}
\end{equation}

\noindent in our list-mode data, where LG is the long gate integral and SG is the short gate integral described in Fig.~\ref{fig:custom_PSD_diagram}. From our PSD plots in Fig.~\ref{fig:PSD}, we chose a constant tail-to-total integral value to discriminate between the neutron (higher R) and photon (lower R) bands spanning over all light-output values.
In practice, we discriminate separated light output slices and fit a curve to chosen separation points.\cite{polack_algorithm_2015} For simplicity in this analysis, we chose constant R values representing the minimum observed overlap at low light output. 
R values of 0.24, 0.27, and 0.24 were selected for the $^{252}$Cf, AmLi, and PuBe sources, respectively. 
Using each respective discrimination value, we divided the neutron source list format data into discriminated neutron and photon data for spectra plotting and multiplicity analysis. The PSD is applied after the measurement is completed. To evaluate our PSD fidelity visually during measurement, we utilized the CoMPASS on-the-fly PSD plotting feature.

\subsection{Signal-triggered coincidence gates}
\label{subsec:sigtrig}
\begin{figure}[h!]
    \centering
    \includegraphics[trim= 90 330 60 200,clip,width=0.6\textwidth]{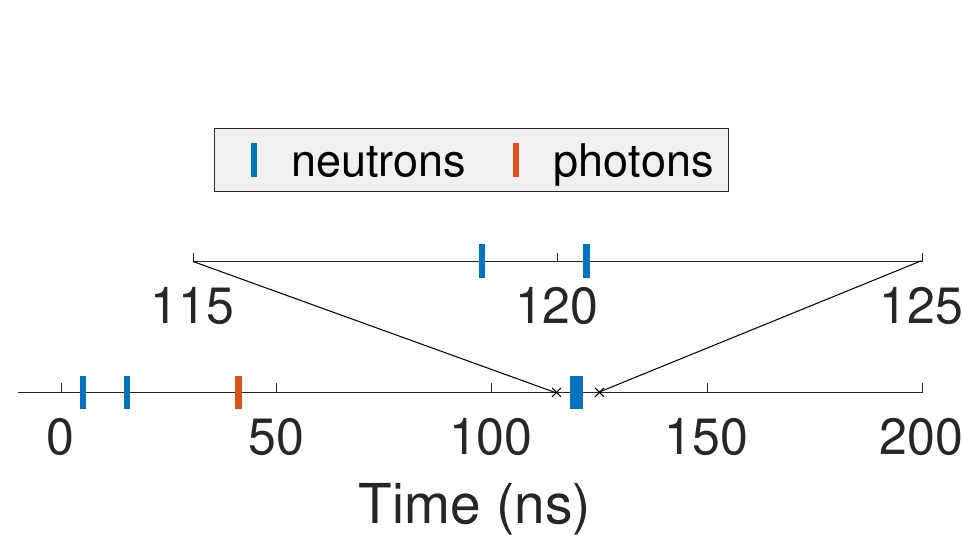}
    \caption{(color online) One 200 ns pulse train of particle detections that were measured for the $^{252}$Cf source with an expanded coincident subset.}
    \label{fig:train}
\end{figure}

We tagged each detection in our list format data with the appropriate particle label after discriminating the detections into neutron and photon events and sorted the events by time of detection. 
All detections were treated as gate triggers starting from the first detected particle and a 200 ns gate was opened forward in time after each trigger. 
The combination of neutron and photon detections in each gate was recorded and included the triggering signal. 
Each unique combination was summed regardless of the order so that a combination of the order ``NP'' is equivalent to ``PN'' where ``N'' is a neutron and ``P'' is a photon detected in the same gate. 
All possible combinations of four or fewer particles were tallied separately and instances of five or more detections in a gate were tallied without recording the combination type. 
The method described above is signal-triggered multiplicity counting and is used to create our combined neutron and photon multiplicity distribution. 
Figure~\ref{fig:train} in combination with Table~\ref{tab:train} provides one example of a pulse train and the signal-triggered combinations that would be recorded for these pulse trains. 
This pulse train shows a combination of four neutrons and one photon recorded during the $^{252}$Cf measurement. 
These detection timelines of multiple coincident neutrons and photons are limited to a maximum of one detection of each unique combination. 
For example, the timeline in Fig.~\ref{fig:train} would record one ``$>$4'', one ``NNNP'', one ``NNP'', one ``NN'', and one ``N'' combination.
One combination for each trigger forward in time.

\begin{table}[h!]
    \centering
    \caption{Signal-triggered combinations that would be added to the mixed multiplicity distribution from the pulse train in Fig.~\ref{fig:train} with 200 ns signal-triggered gates. }
    \label{tab:train}
    \begin{tabular}{cl|c|c|c|c|c|c|c|c|c|c|c|c|c|c|c}
        & \textbf{Detection timeline} & \rot{N} & \rot{P} & \rot{NN} & \rot{\textit{NP}} 
        & \rot{PP} & \rot{NNN} & \rot{NNP} 
        & \rot{NPP}& \rot{PPP} & 
        \rot{NNNN} & \rot{NNNP} & \rot{NNPP} & \rot{NPPP} & \rot{PPPP} & \rot{$>4$} \\
        \hline
        & $^{252}$Cf data in Figure~\ref{fig:train}               & \OK &   & \OK  &   &   &   & \OK &   &  &  & \OK &  &  &  & \OK \\
    \end{tabular}
\end{table}


\section{Results and Discussion}

\subsection{Neutron source light-output spectra}

\begin{figure}[h!]
    \centering
    \begin{subfigure}{0.99\textwidth}
        \includegraphics[width=0.8\textwidth]{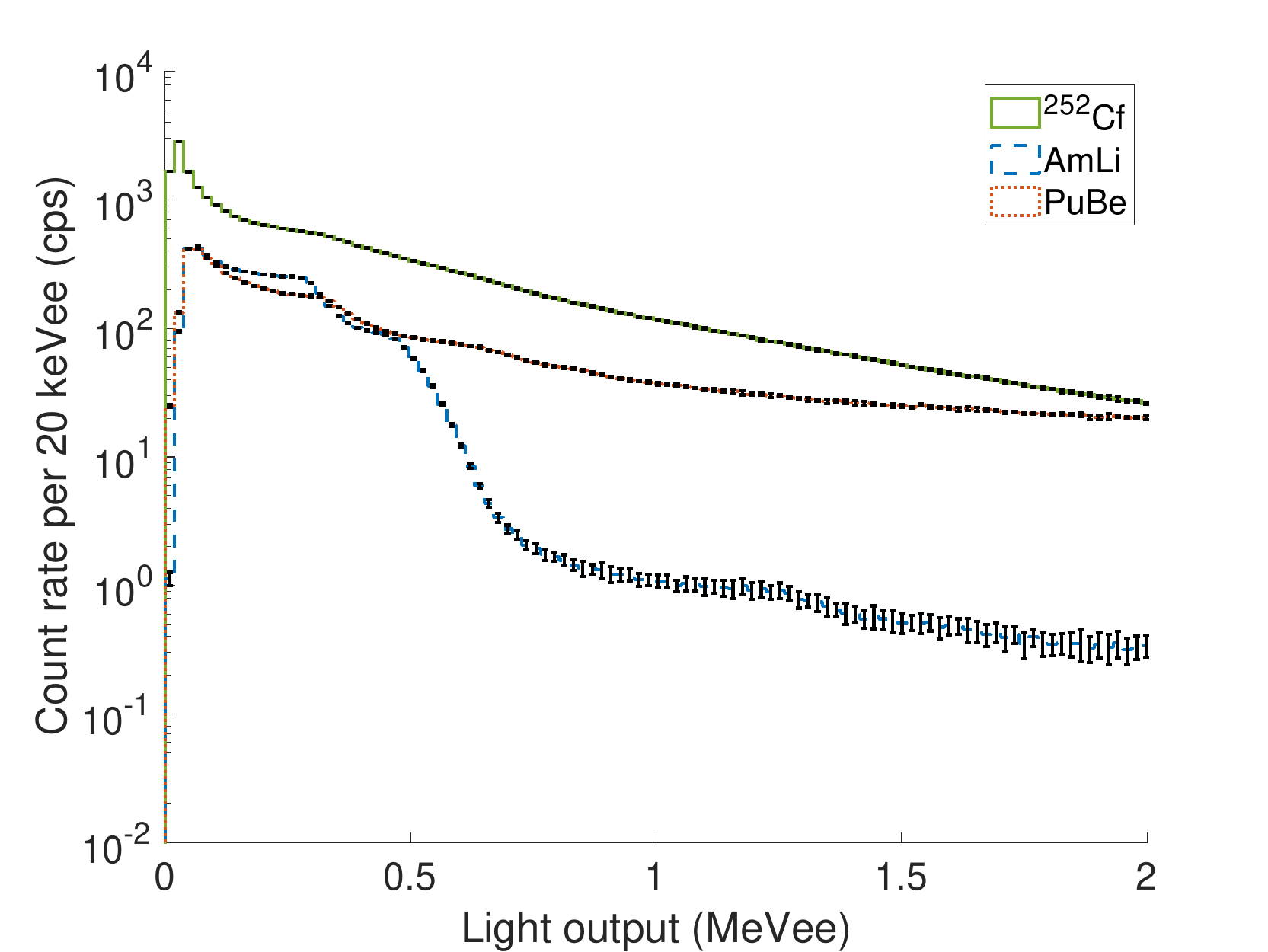}
        \caption{}
        \label{fig:photon_LO_specs}
    \end{subfigure}
    \begin{subfigure}{0.99\textwidth}
        \includegraphics[width=0.8\textwidth]{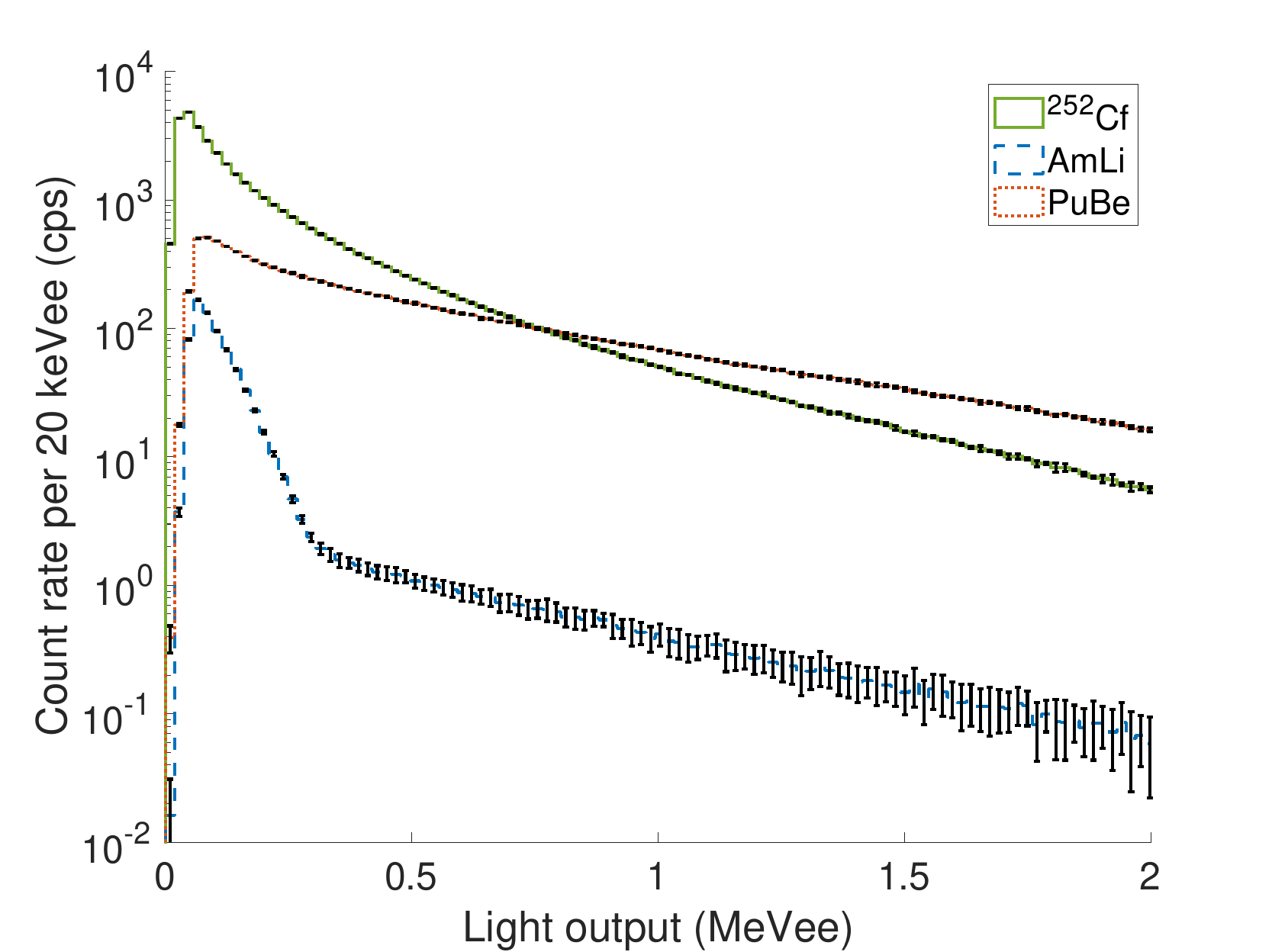}
        \caption{}
        \label{fig:neutron_LO_specs}
    \end{subfigure}
    \caption{(color online) Total light-output spectra for each neutron source for (a) photons and (b) neutrons.}
    \label{fig:LO_specs}
\end{figure}

The light-output spectra in Fig.~\ref{fig:LO_specs} compare the absolute photon distributions in count rate per bin for each neutron source (\ref{fig:photon_LO_specs}) and the absolute neutron distributions in count rate per bin for each neutron source (\ref{fig:neutron_LO_specs}) based on the PSD in the previous section. 

The AmLi source has approximate observed maximum energy scatters at about 0.80 MeVee and 0.31 MeVee (the latter due to neutron light-output quenching) on the photon and neutron light-output spectra, respectively. 
The light output above these values is likely due to multiple scatter events from a single particle (in the photon spectrum) or pulse pile-up (in both spectra). 
Observable maximum energy scatters for AmLi distinguish it from a $^{252}$Cf fission source and show that the energy sensitivity of our organic scintillator system provides additional capability as compared to $^3$He systems given their lack of direct energy sensitivity.
The spectra do not fully distinguish PuBe from $^{252}$Cf so further multiplicity analysis is required.

\subsection{Combined neutron and photon multiplicity of neutron sources}

\begin{figure}[ht]
    \centering
    \begin{subfigure}{0.69\textwidth}
        \includegraphics[width=\textwidth]{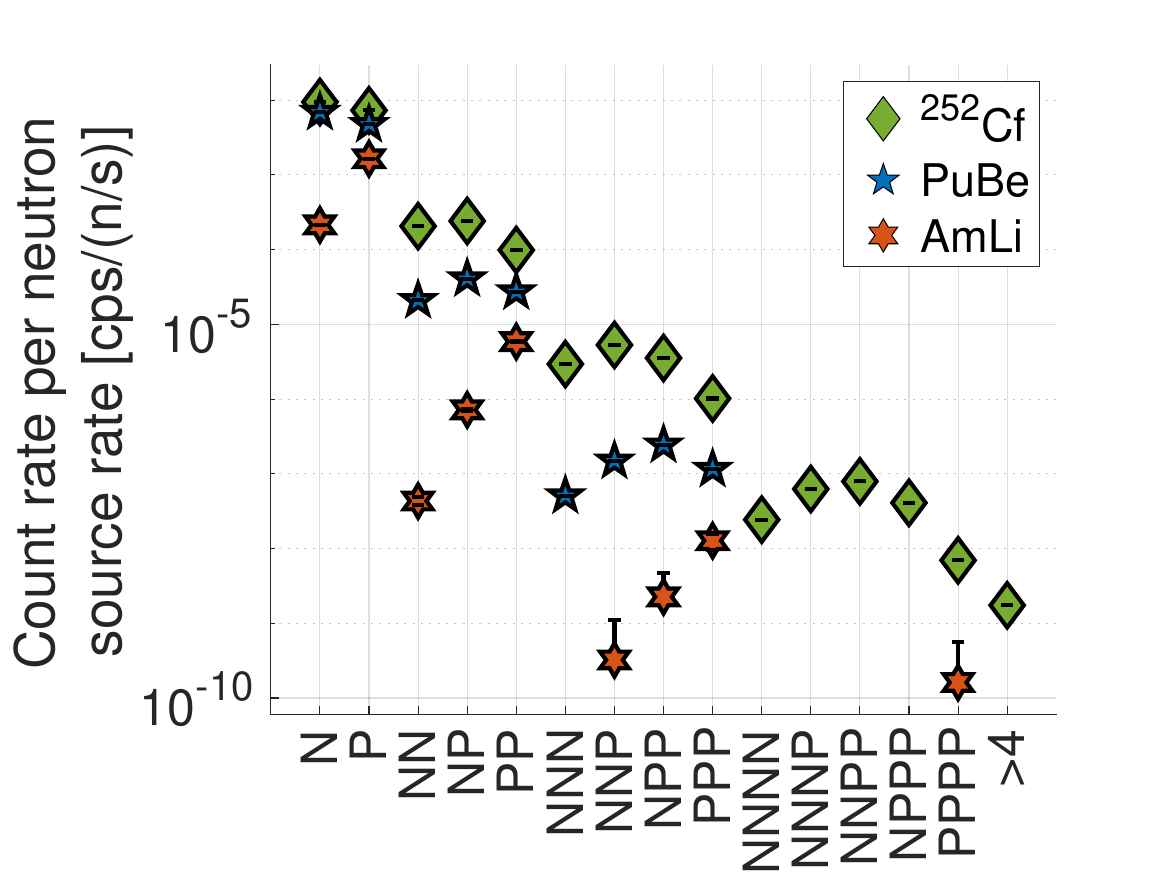}
        \caption{}
        \label{fig:mm}
    \end{subfigure}
    \begin{subfigure}{0.69\textwidth}
        \includegraphics[width=\textwidth]{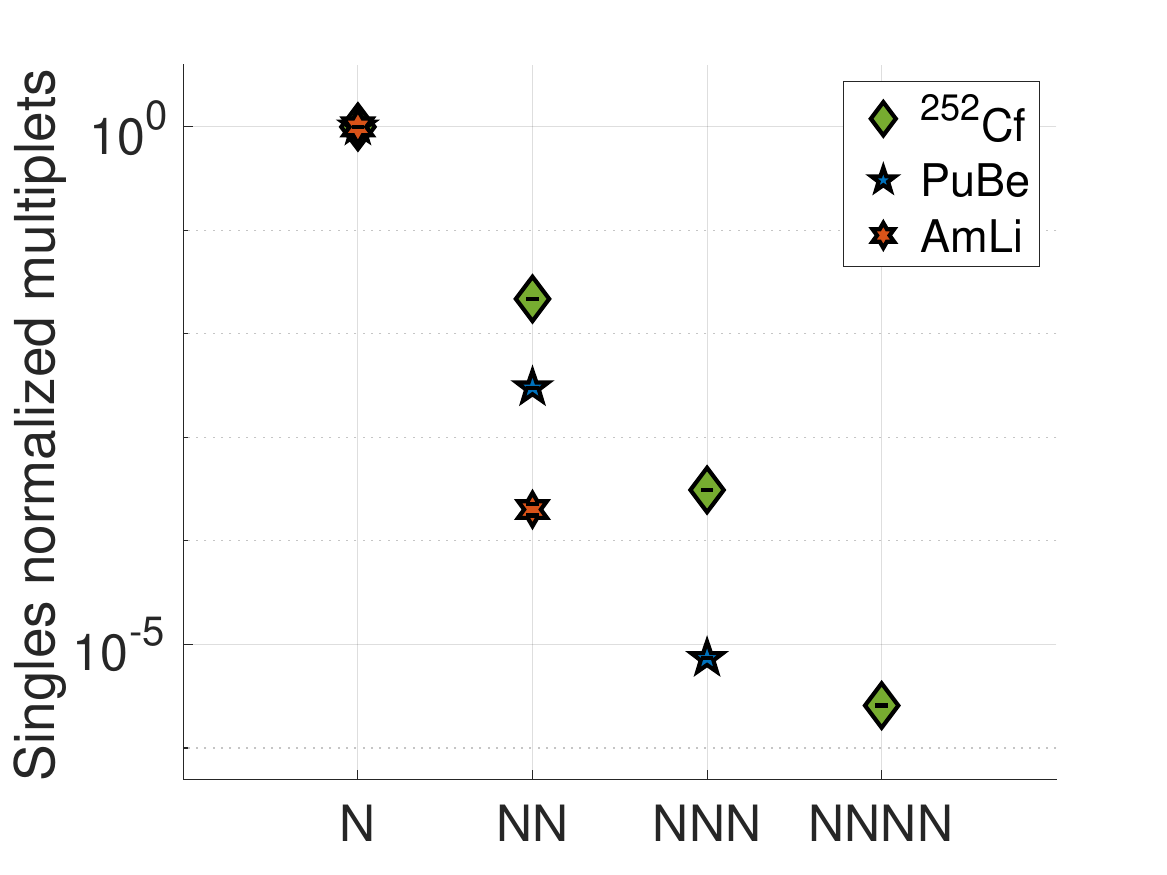}
        \caption{}
        \label{fig:nm}
    \end{subfigure}
    \caption{(color online) The mixed multiplet combinations\cite{di_fulvio_fast_2015} of photon and neutron detections (a) as count rates normalized by neutron source rate of each source and (b) neutron multiplets normalized by the neutron singles rate for each source. When only upper error bars are shown, the mean value minus one standard deviation is less than zero.}
    \label{fig:multiplet_plots}
\end{figure}

We obtained the combined neutron and photon multiplicity distribution of each neutron source per unit neutron source strength in Fig.~\ref{fig:mm} by applying a 200 ns, signal-triggered timing gate to the discriminated list format data from each radiation source as described in Sec.~\ref{subsec:sigtrig}.
The 0.035 MeVee threshold applied is equivalent to about a 0.350 MeV neutron scatter on a proton in stilbene, which eliminates more of the total neutron scatters for AmLi than PuBe or $^{252}$Cf due to the higher proportion of neutron scatter events above the energy threshold for the latter two sources.
Consequently, the AmLi exhibits a much lower single neutron counts per source neutron than $^{252}$Cf and PuBe due to the lower relative neutron energies emitted from AmLi. 
The AmLi ``NNP'', ``NPP'', and ``PPPP'' combinations are all within one standard deviation of zero, and all combinations that are not shown are zero.
The AmLi does not produce any neutron triples or particle combinations of four or more, further distinguishing AmLi from $^{252}$Cf. 
The $^{252}$Cf produces neutron quadruples and particle combinations of five or more, which is reflective of the higher multiplicity of $^{252}$Cf. 
PuBe produced all combinations of triples or less, but at a significantly lower magnitude than $^{252}$Cf despite a similar neutron and photon source-rate-normalized singles magnitude. 
PuBe did not produce any combinations of four particles except for ``PPPP'' events that were within one standard deviation of zero. 
The PuBe source produced significantly more neutron double counts than the AmLi source and the PuBe source had measurable triple neutron counts. 
These points indicate that PuBe may have a multiplicity, but it is much lower than that of the $^{252}$Cf.

The neutron multiplet rates normalized by the neutron singles rate in Fig.~\ref{fig:nm} show that the relative multiplets of the $^{252}$Cf source are clearly higher than that of the ($\alpha$,n) sources.
The small rates of neutron multiplets relative to the neutron singles from AmLi completely distinguish AmLi from $^{252}$Cf  and PuBe. Neutron cross talk contributes to the small rate of neutron doubles from AmLi. The neutron doubles and rare neutron triples relative to the neutron singles from PuBe arise from several characteristics of the PuBe source. The PuBe has a higher proportion of neutrons above the detection threshold than both $^{252}$Cf and AmLi, leading to a higher likelihood of neutron cross talk. Secondary (n,2n) reactions in the PuBe matrix, trace amounts of $^{240}$Pu, and (n, fission) reactions increase the detected multiplets as well. However, the relative multiplet rates from PuBe are sufficiently low to distinguish the PuBe from $^{252}$Cf.

\section{Conclusions}

We have developed an advanced instructional laboratory that teaches students how to operate a complex organic scintillator well counter and apply nondestructive assay techniques. 
By completing this laboratory, students will be capable of measuring neutron sources with organic scintillators using the CoMPASS software and CAEN electronics. 
Participants will learn about pulse-shape discrimination concepts and techniques and be capable of producing discriminated light-output spectra. 
Building on these methods with signal-triggered time gating, students will be able to produce combined neutron and photon multiplicity distributions. 
$^{241}$Am-Li (AmLi) can be distinguished from $^{252}$Cf using the relative maximum energy scatters of neutron and photon light-output spectra and multiplicity distribution relative to $^{252}$Cf. 
In addition, $^{239}$Pu-Be (PuBe) can be distinguished from the multiplicity distribution relative to $^{252}$Cf with some $^{240}$Pu. 
The difference in relative AmLi and PuBe multiplets may also provide a signature for the trace amounts of plutonium, a special nuclear material, in PuBe. 
Further simulation of cross talk is required to fully rule out an ($\alpha$,n) source as a fission source, but neither ($\alpha$,n) has spoofed the $^{252}$Cf fission source as a non-fissioning neutron source due to the analysis methods applied. 
This laboratory builds all the fundamental skills needed to assay special nuclear materials like plutonium and uranium in conjunction with gamma-ray spectroscopy for material isotopics\cite{reilly_passive_1991,sampson_plutonium_1980} and the cross-talk adjusted point-model equations for neutron multiplet evaluation.\cite{shin_neutron_2017}


\begin{acknowledgments}

We would like to thank the Detection and Nuclear Nonproliferation Group at the University of Michigan for supporting this work and for the development of the fast neutron multiplicity well-counter.
We would like to especially thank Professor Angela Di Fulvio and Dr. Tony Shin for their work in developing and implementing this course with EJ-309 liquid scintillators and the trans-stilbene organic scintillators used in this laboratory.
This work is supported and funded by the Consortium for Monitoring, Technology, and Verification under DOE NNSA award number DENA0003920. 
Any opinions, findings, conclusions, or recommendations expressed in this material are those of the authors and do not necessarily reflect the views of the Consortium for Monitoring, Technology, and Verification.

\end{acknowledgments}

\section{Author Declarations}

\subsection{Pre-proof Statement}
The authors are submitting this work to ArXiv in its pre-proof form.
This article is being processed for publication by the American Institute of Physics for publication in the American Journal of Physics in the Instruction Laboratories and Demonstrations subsection.
The processing identification for this pre-proof is Article Number: AJP22-AR-01524.
AJP has provided permission to the authors to submit this pre-proof to ArXiv.

\subsection{Conflict of Interest Disclosure}
The authors have no conflicts to disclose.

\section{References}

\bibliography{main.bib}  
\bibliographystyle{ieeetran}

\appendix   

\section{Possible Extensions and Cost-saving Adjustments}

The mixed multiplicity laboratory utilizes several unique and complex characteristics of organic scintillators to provide an advanced teaching laboratory. 
Calculating discriminated light-output spectra and mixed multiplicity distributions is a challenging task, but there are other aspects of this laboratory that may be separately focused on or added in addition to this experiment presented to perform useful instructional laboratories.

Pulse-shape discrimination (PSD) optimization is a challenging parameter optimization problem that could be explored with only one detector and neutron source.\cite{polack_algorithm_2015}
The confidence in PSD is a function of the tail-to-total integral ratio distribution within a given light-output range.
This confidence can be evaluated by a figure of merit (FOM) such as:

\begin{equation}
    \label{PSD:FOM}
    \text{FOM} = \frac{\mu_n-\mu_p}{\text{FWHM}_n + \text{FWHM}_p},
\end{equation}

\noindent where $\mu_n$ and $\mu_p$ are the mean value of Gaussian fits to the neutron and photon ratio distributions and $\text{FWHM}_n$ and $\text{FWHM}_p$ are the full-width at half-maximum of each respective Gaussian fit. 
By optimizing for the best FOM in the desired light-output range, the short gate can be optimized for any organic scintillator.
Students would learn the underlying physics behind delayed fluorescence,\cite{hamel_plastic_2021} curve-fitting for histogram distributions, and figure-of-merit evaluation.

A neutron time-of-flight (TOF) laboratory may be completed with only two detectors from our setup, one neutron source, and the optional addition of a shadow bar to account for room return.\cite{becchetti_cf-252_2013} 
The neutron light-output response relative to the photon light-output response of organic scintillators is nonlinear with energy, which makes it difficult to calculate incident neutron energy.
TOF measurements provide the capability to calculate the neutron energy spectrum from a neutron source with organic scintillators using coincident signals from a trigger and offset detector.
TOF also provides a means of quantifying PSD particle misclassification.
Placing a $^{252}$Cf adjacent to a trigger detector and offset from a coincident organic scintillator the neutron energy can be derived from the neutron time of flight and classical physics.
The shadow bar is composed of mainly hydrogen\cite{hopkins_shadow-bar_1967} and attenuates neutrons along the direct flight path between two detectors.
By subtracting the shadow bar TOF from the unattenuated TOF, room return is accounted for. 
Students would learn the $^{252}$Cf energy spectrum and optimization of detector offset for time-of-flight resolution and detection efficiency.

Similarly, organic scintillators may uniquely observe the cross-correlation of fission events\cite{enqvist_measurement_2008,di_fulvio_fast_2015,marcath_plutonium_2014} with only two detectors and one fission source. 
Placing a $^{252}$Cf equidistant between two coincident organic scintillators the symmetry and dimensions of the measurement system may be inferred from peaks in the discriminated neutron (n) and photon (p) coincident time-difference distributions. 
Additionally, particle misclassification may be estimated from the overlap of the respective (n,n), (p,p), and (n,p) cross-correlation distributions. 
Students would apply discriminated coincident time-difference calculations to build a cross-correlation distribution and classical physics for neutron and photon time of flight.

Fission events have anisotropy in particle emission\cite{skarsvag_prompt_1977,holewa_using_2013} and the well-counter utilized in this laboratory is well-equipped to measure coincidence anisotropy.\cite{shin_neutron-neutron_2019} The anisotropy can be quantified as a coincidence ratio of respective detector pair groups at specified angles. 
Students would learn how to complete advanced coincidence analysis of large detector arrays.

The equipment used for this laboratory may be substituted for lower-cost alternatives. 
EJ-309 liquid scintillators as well as plastic scintillators are capable organic scintillators for multiplicity counting,\cite{di_fulvio_passive_2017,cohen_demonstration_2022} but do not have the same low-energy PSD capability as trans-stilbene crystals.
Higher measurement light-output thresholds would be necessary to reduce particle misclassification.
Higher measurement thresholds require longer measurement times to achieve the same statistics, which would increase the measurement time required for this laboratory.
A promising new, low-cost alternative that remains untested for multiplicity counting is organic glass.\cite{shin_measured_2019} The PSD of organic glass is not as good as trans-stilbene, but organic glass has comparable timing precision to trans-stilbene and comparable PSD to EJ-309.
The photomultiplier tubes may be replaced by silicon photomultipliers (SiPMs)\cite{sabet_high-performance_2012,chung_development_2022} at a potentially lower cost.
The digitizer\cite{skulski_FemtoDAQ_2017} and high voltage units may also be replaced by lower cost alternatives at the drawback of decreased ease-of-use and loss of built-in compatibility with routinely maintained software like CoMPASS.
The digitizer could be swapped for a cheaper, lower frequency (less than 500 MHz) CAEN unit at a loss of PSD fidelity and pulse timing precision.


\end{document}